\documentclass[prd,twocolumn,showpacs,preprintnumbers,amsmath,amssymb,floatfix]{revtex4}

\usepackage{graphicx,epsfig}
\usepackage{dcolumn}
\usepackage{bm}

\def\lsim{\mathrel{\lower2.5pt\vbox{\lineskip=0pt\baselineskip=0pt
\hbox{$<$}\hbox{$\sim$}}}}
\def\gsim{\mathrel{\lower2.5pt\vbox{\lineskip=0pt\baselineskip=0pt
\hbox{$>$}\hbox{$\sim$}}}}

\newcommand{\be}{\begin{equation}}
\newcommand{\ee}{\end{equation}}

\newcommand{\Ls}{\Lambda(1405)}
\newcommand{\ba}{\begin{eqnarray}}
\newcommand{\ea}{\end{eqnarray}}

\begin{document}

\preprint{TUM/T39-07-25, YITP-07-97}

\title{Structure of the $\Lambda(1405)$ baryon resonance 
from its large $N_c$ behavior}

%
%

\author{
Tetsuo~Hyodo,$^{1,2}$
Daisuke~Jido,$^{2}$ and 
Luis~Roca$^{2,3}$}

\affiliation{
$^1$ Physik-Department, Technische Universit\"at M\"unchen, 
D-85747 Garching, Germany \\
$^2$Yukawa Institute for Theoretical Physics, Kyoto University, Kyoto 606--8502, Japan \\
$^3$Departamento de F\'{\i}sica. Universidad de Murcia. E-30071, Murcia. Spain
}

\date{\today}

\begin{abstract}

We study the behavior with the number of colors ($N_c$)  of the two poles 
associated to the $\Lambda(1405)$ resonance obtained dynamically within the
chiral unitary approach. The leading order chiral meson-baryon interaction 
manifests a nontrivial $N_c$ dependence for SU(3) baryons, which gives a 
finite attractive interaction in some channels in the large $N_c$ limit. As a
consequence, the SU(3) singlet ($\bar{K}N$) component of the $\Lambda(1405)$ 
survives in the large $N_c$ limit as a bound state, while the other 
components dissolve into the continuum. The $N_c$ dependence of the decay 
widths shows different behavior from the general counting rule for a $qqq$ 
state, indicating the dynamical origin of the two poles for the 
$\Lambda(1405)$  resonance.

\end{abstract}


\maketitle


The quest for the understanding of the nature of the hadronic spectrum is one
of the most challenging issues in particle physics. Especially the structure 
of hadronic resonances is one of the cornerstones in this field. Recently it 
has been suggested that multiquark and/or hadronic components of some 
resonances in intermediate energies prevail over the simple mesonic $q\bar q$
and  baryonic $qqq$ components. For instance, the light scalar mesons have 
been investigated in the four-quark picture~\cite{Jaffe:1976ig}, in the 
mesonic molecular picture~\cite{Oller}, and in lattice 
QCD~\cite{mesonlattice}. Methods to clarify the internal structure of the 
hadrons are called for. This is one of the main aims of the present work, in 
the baryonic sector.

Despite the success of QCD as the theory for the strong interactions, the 
inherent confinement properties of quarks and gluons make its direct 
applicability to investigate the hadron structure far from being feasible. 
One of the most powerful tools to investigate the properties of hadrons are 
the effective theories of QCD, such as chiral perturbation theory 
(ChPT)~\cite{chpt} and its recent development, chiral unitary 
approach~\cite{Oller,KaiserOset,Oller:2000fj,Lutz:2001yb}. ChPT provides 
hadronic interactions based on a systematic low-energy expansion. The 
nonperturbative resummation of the low-energy interactions required by the 
unitarity condition leads to the success of the chiral unitary approaches in 
a variety of hadron scatterings with many hadronic resonances being 
dynamically generated.

It is particularly interesting that the $\Lambda(1405)$ is well reproduced in
the chiral unitary approach as a dynamically generated resonance in the 
$s$-wave meson-baryon scattering with strangeness $S=-1$ and isospin 
$I=0$~\cite{KaiserOset,Oller:2000fj,Lutz:2001yb}. This finding unveils the 
structure of the $\Lambda(1405)$ as a superposition of two 
states~\cite{Oller:2000fj,Jido:2003cb}, which have been studied both 
theoretically and experimentally~\cite{test_twopoles}. The structure of the 
$\Lambda(1405)$ has a large impact for the study of the $\bar{K}N$ 
interaction~\cite{Hyodo:2007jq}, which is eventually important for the study 
of the kaonic nuclei and kaon condensation in the neutron 
star~\cite{Kaplan:1986yq}. The works based on the chiral unitary approach 
have shed new light on the understanding of the nature of the $\Ls$ 
resonance, reinforcing the dynamical or ``molecular" origin versus the $qqq$ 
picture.

The expansion in the inverse of the number of colors, $1/N_{c}$, is an 
analytic, well established and widely used approximation to QCD valid for the
whole energy region, and enables us to investigate the qualitative features 
of hadrons~\cite{Ncscaling}. In past years the dependence on $N_c$ of the 
resonance properties within the chiral unitary approach has shown up as a 
powerful tool to discriminate the quark structure for particular mesonic 
resonances~\cite{largeNCmesons,newaxialsNc}. The study of the $N_c$ scaling 
allows for a clear identification of $q\bar q$ mesonic states based on the 
general scaling properties of their masses as ${\cal O}(1)$ and the widths as
${\cal O}(1/N_c)$~\cite{Ncscaling}. These $N_c$ behaviors were compared to 
the chiral unitary approach predictions in order to clarify the internal 
structure of the resonances. Recent interest in the large $N_{c}$ argument of
low-energy hadrons is also related to the holographic 
QCD~\cite{SakaiSugimoto}.
 
In this paper, we present the first study of the $N_c$ behavior of the 
physical baryon resonances in the chiral unitary approach, focusing on the 
structure of the $\Lambda(1405)$. The study of the baryon  resonances has an
important difference from the meson sector: the nontrivial $N_c$ dependence 
of the leading order chiral interaction~\cite{hyodo}. The $N_c$ dependence 
stems from the change of the flavor  representation of the baryons with 
$N_c$, when the number of flavors is larger than 2~\cite{changeflavor}. As 
a consequence, the meson-baryon interactions for some channels remain finite 
in the large $N_c$ limit, leading to nontrivial consequences for the 
generated resonances. The study of the $N_c$ behavior also tells us about the
quark structure of the generated resonances. The general $N_c$ counting rule 
for ordinary $qqq$ baryons indicates the scaling of the decay width as 
$\Gamma\sim {\cal O}(1)$, the mass  $M_R~\sim {\cal O}(N_c)$ and the 
excitation energy $\Delta E\sim{\cal O}(1)$ \cite{Goity_cohen}. Hence any 
significant deviation from these behaviors indicates that the molecular, or
dynamical, component of the $\Ls$ resonance dominates over the $qqq$ 
contribution. The $N_c$ behavior of the baryon resonances has been studied in
Ref.~\cite{Garcia-Recio:2006wb} in the SU(6) symmetric limit. Here we 
consider physical $\Lambda(1405)$ resonance including the flavor symmetry
breaking effects.

The $\Lambda(1405)$ in the chiral unitary approach is dynamically generated 
in the $s$-wave meson-baryon scattering with $S=-1$ and $I=0$. Based on the 
N/D method~\cite{Oller:2000fj}, the coupled-channel scattering amplitude 
$T_{ij}$ is given by the matrix equation
\begin{equation}
   T=[1-VG]^{-1}V ,
   \label{eq:BS}
\end{equation}
where $V_{ij}$ is the interaction kernel and the function $G_{i}$ is given 
by the dispersion integral of the two-body phase space 
$\rho_i(s)=2M_{i}\sqrt{(s-s_i^+)(s-s_i^-)}/(8\pi s)$ in a diagonal matrix 
form by
\begin{align}
   G_i(W)
   &=-\tilde{a}(s_0)
   -\frac{s-s_0}{2\pi}
   \int_{s_i^{+}}^{\infty}ds^{\prime}
   \frac{\rho_i(s^{\prime})}{(s^{\prime}-s)(s^{\prime}-s_0)}
   \label{eq:loop_s} , \\
   &s=W^{2} , \quad s_i^{\pm}=(m_i\pm M_{i})^2,
   \nonumber
\end{align}  
where $W$ is the center-of-mass energy, $s_0$ the subtraction point,
$\tilde{a}_i(s_0)$ the subtraction constant, and $M_i$ ($m_i$) is the mass
of the baryon (meson) of the channel $i$. The degree of freedom introduced by
the subtraction constant is equivalent to the one from the cutoff of the loop
integral in the scattering equation, and is fixed following the prescription 
given in Refs.~\cite{Lutz:2001yb,hyodo}. Namely, we impose the condition 
$G_i(\mu) = 0$, where $\mu$ is the matching scale of the full amplitude 
$T_{ij}$ to the interaction kernel $V_{ij}$ and we take $\mu = M_i$. For the 
physical scattering case, the scale $\mu$ is taken to be $M_{\Lambda}$. We 
will also examine the three-momentum cutoff scheme later.

The interaction kernel $V_{ij}(W)$ in Eq.~\eqref{eq:BS} is the driving force
to generate the resonances. The leading order term of ChPT provides the 
$s$-wave meson-baryon interaction for energy $W$ as
\begin{align}
  V_{ij}(W)=&-C_{ij}\frac{1}{4f^2}(2W-M_i-M_j)\,\eta_i\,\eta_j,
  \label{eq:WT}
\end{align} 
where $\eta_l=\sqrt{(M_l+E_l)/(2M_l)}$, $C_{ij}$ expresses the coupling 
strength, $f$ is the pseudoscalar meson decay constant, and $E_i$ is the 
energy of the baryon in channel $i$. Equation~\eqref{eq:WT} is known as the 
Weinberg-Tomozawa (WT) term derived using current algebra~\cite{WT}. 

The $N_c$ dependence of the parameters is introduced as $M_i\propto N_c$, 
$m_i\propto 1$, and $f\propto \sqrt{N_c}$~\cite{Manohar:1998xv}. In the 
following, we will discuss the $N_c$ dependence in the coupling strengths 
$C_{ij}$ in SU(3) and isospin basis.

Let us first consider a simple case in the SU(3) symmetric limit. The 
coupling strengths $C_{ij}$ are determined by group theoretical arguments,
including the $N_c$ dependence~\cite{hyodo}. In the SU(3) basis, the relevant
coefficients for $S=-1$ and $I=0$ are given by the diagonal matrix
\be
C_{ij}^{SU(3)}(N_c)
=\textrm{diag}\left(\dfrac{9+N_c}{2},3, 3,
\dfrac{-1-N_c}{2}\right),
\label{eq:couplingSU3Nc}
\ee
with the channels being $\bm{1}$, $\bm{8}$, $\bm{8}'$, and $\bm{27}$. At 
$N_c=3$, we find attractive interaction for the singlet and the two octet 
channels. It has commonly been considered that the WT term~\eqref{eq:WT}
behaves as $1/N_c$ because of the $1/f^2$ factor~\cite{Manohar:1998xv}. Here 
we would like to emphasize that, in the baryon case, the linear $N_{c}$ 
dependence on the coupling strength $C$ indicates an ${\cal O}(1)$ attractive
(repulsive) interaction in the singlet (27-plet) channel even at the large 
$N_c$ limit. This is not contradictory to the general counting rule of 
meson-baryon scattering~\cite{Ncscaling,Manohar:1998xv}.

The derivation of Eq.~\eqref{eq:couplingSU3Nc} is based on the standard $N_c$
extension for the baryon, that is, the irreducible representation $[p,q]$ of 
the flavor SU(3) at $N_{c}=3$ is extended to $[p,q+(N_c-3)/2]$ for an 
arbitrary $N_{c}$. In this prescription, the spin, isospin, and strangeness 
of
the baryon are the same with those at $N_c=3$, while the baryon has different
charge and hypercharge from those at $N_{c}=3$. Here we adopt this standard 
$N_{c}$ extension, since it is convenient for the flavor SU(3) breaking. 
There are two more extensions~\cite{changeflavor}; 
$[p,q] \to [p+(N_c-3)/3,q+(N_c-3)/3]$ and 
$[p,q] \to [p+N_c-3,q]$. These extensions have some advantages, but the
baryons constructed in these ways have unphysical strangeness and spin. We 
have confirmed that it is also the case in these $N_c$ extensions that the 
singlet (27-plet) channel has positive (negative) $N_c$ dependence as seen 
in Eq.~\eqref{eq:couplingSU3Nc}.

Since  in the SU(3) basis the scattering equation~\eqref{eq:BS} is a set of 
single-channel problems, the existence of bound states can be studied by 
comparing the coupling strengths~\eqref{eq:couplingSU3Nc} with the critical 
coupling $C_{\text{crit}}$ introduced in Refs.~\cite{hyodo}:
\begin{equation}
   C_{\text{crit}}(N_c)= \frac{2[f(N_c)]^2 
   }{m_i[-G(M_{i}(N_c)+m_i)]} .
   \label{eq:Ccrit}
\end{equation}
If the coupling strength of the channel $i$ is larger than this critical
value, a bound state is generated. Studying the $N_c$ dependence, we find 
that $C_{\text{crit}}(N_c)$ increases more slowly than the $N_c/2$ growth of
the coupling  strength in the singlet channel. This means that the bound 
state at $N_{c}=3$ in the singlet channel~\cite{Jido:2003cb} still survives 
in the large $N_{c}$ limit. This is a nontrivial consequence of the $N_c$ 
dependence of the WT interaction for the baryon. On the other hand, the 
attraction in the octet channels becomes smaller than $C_{\text{crit}}(N_c)$ 
at larger values of $N_c$ and the bound states will disappear in the large 
$N_c$ limit. The $N_c$ dependence of the coupling strengths given in 
Ref.~\cite{hyodo} shows that $\Lambda(1405)$ is the only example of hadron 
bound states of $qqq$ baryon and the Nambu-Goldstone boson in the large $N_c$
limit. 

\begin{figure}[bp]
   \centering
   \includegraphics[width=.5\textwidth,clip]{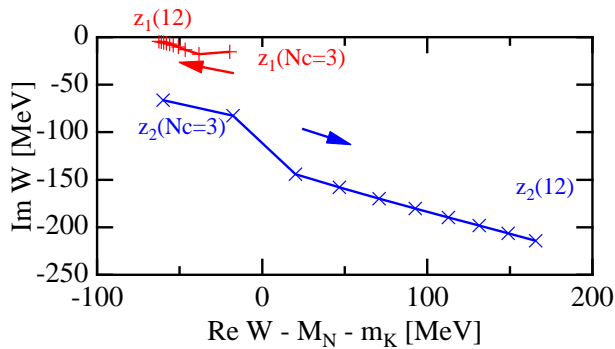}
   \caption{(Color online) Pole positions for the $\Lambda(1405)$ in 
   the complex energy plane as functions of integer $N_c$ from $N_{c}=3$ to
   $12$.}
   \label{fig:poles}
\end{figure}%

Now we turn to the central problem of the physical $\Lambda(1405)$. We 
introduce SU(3) breaking in the masses of the particles to discuss physical 
$\Lambda(1405)$ and work in the isospin basis. The coupling 
strengths~\eqref{eq:couplingSU3Nc} can be translated into the isospin basis 
by the SU(3) Clebsch-Gordan coefficients with $N_c$ 
dependence~\cite{Cohen:2004ki} as
\begin{align}
    C_{ij}^I(N_c)
    &=
    \begin{pmatrix}
        \frac{{N_c}+3}{2}
	& -\frac{\sqrt{3N_c-3}}{2} 
	& \frac{\sqrt{3N_c+9}}{2}
	& 0   \\
	& 4
	& 0
	& \frac{\sqrt{N_c+3}}{2} \\ 
	& 
	& 0
	& -\frac{\sqrt{9N_c-9}}{2} \\
	& 
	& 
	&  \frac{-{N_c}+9}{2} \\
     \end{pmatrix}
    \label{eq:couplingINc}
\end{align}
with the channels being $\bar{K}N$, $\pi\Sigma$, $\eta\Lambda$, and $K\Xi$.

The strengths in diagonal $\bar{K}N$ and $K\Xi$ channels are 
$\mathcal{O}(N_c)$, and the negative $N_c$ dependence in the $K\Xi$ channel 
changes the sign of the interaction from attractive to repulsive for $N_c>9$.
On the other hand, the off-diagonal elements (and diagonal $\pi\Sigma$ and 
$\eta\Lambda$ ones) are $\mathcal{O}(\sqrt{N_c})$ or $\mathcal{O}(1)$, so 
that any transitions among these channels vanish either as $1/\sqrt{N_c}$ or 
$1/N_c$. This means that the meson-baryon scattering in this sector becomes 
essentially a set of single-channel problems in the large $N_c$ limit, even 
with the SU(3) breaking. The coupling strength of $\bar{K}N$ channel in the 
large $N_c$ limit is the same as the one of the singlet channel in the SU(3) 
basis. Therefore, by following the same argument as in the SU(3) symmetric
case, we conclude that there is one bound state in the $\bar{K}N$ channel in 
the large $N_{c}$ limit. It was found in Ref.~\cite{Hyodo:2007jq} that the 
$\bar{K}N$ interaction develops a bound state at $N_c=3$, when the transition
to other channels is switched off. Thus, as in the SU(3) singlet channel, the
$\bar{K}N$ bound state found at $N_c=3$ remains in the large $N_c$ limit in 
contrast to the mesonic resonances, while the other states, such as a 
resonance in $\pi\Sigma$ channel, will disappear.

To study the $N_c$ scaling of the generated resonances, it is interesting to 
observe how the positions of the resonance poles evolve when $N_{c}$ 
increases. At $N_c=3$, we have found two poles associated with 
the physical $\Lambda(1405)$ at $z_1 = -20-15i $ MeV and $z_2 = -61-66i $ 
MeV~\cite{Oller:2000fj,Jido:2003cb}, of which real parts correspond to the 
excitation energy measured 
from the $\bar KN$ channel threshold and the imaginary part expresses the 
half width. We show in Fig.~\ref{fig:poles} the trajectories of the pole 
positions from $N_{c}=3$ to $12$. As $N_{c}$ increases, the pole $z_1$ 
approaches the real axis with reducing the width, while $z_2$ moves to higher
energy region and the imaginary part increases with $N_c$. The substantial 
change of the pole $z_2$ between $N_{c}=4$ and $5$ is due to the fact that 
the pole crosses the $\bar K N$ threshold and hence the width suddenly 
increases because the important $\bar K N$ decay channel opens.

An important finding in Fig.~\ref{fig:poles} is that, as $N_{c}$ increases, 
one resonance tends to become a bound state while the other tends to 
dissipate by moving away from the physical axis. This implies that the width 
of the resonance associated to the pole $z_1$ ($z_2$) decreases (increases) 
as $N_c$ increases, which is in obvious contradiction to the dominance of the
$qqq$ component, whose decay widths should scale as 
$\Gamma\sim \mathcal{O}(1)$~\cite{Goity_cohen}. This observation supports 
the idea of the dynamical origin of the $\Lambda(1405)$ resonance against a 
$qqq$ dominant composition.

As discussed in Ref.~\cite{newaxialsNc}, the cutoff scale may have an $N_c$ 
dependence. We have examined the cutoff regularization scheme for the loop 
function~\eqref{eq:loop_s} with two different $N_c$ dependences of the 
three-momentum cutoff, $q_{\text{max}}\sim 1$ and 
$q_{\text{max}}\sim \sqrt{N_c}$. The results among the different 
regularization schemes are qualitatively similar~\footnote{In some cases, 
$z_2$ is the pole that tends to be a bound state instead of $z_1$. However, 
the study of the pole residues indicates that the dominant components of the 
would-be-bound state is the SU(3) singlet and $\bar{K}N$ state, irrespective 
to the original pole positions at $N_c=3$.}. Hence, the conclusions drawn 
here are independent on the regularization procedure and its $N_c$ 
dependence.  

\begin{figure}[bp]
    \centering
    \includegraphics[width=.49\textwidth]{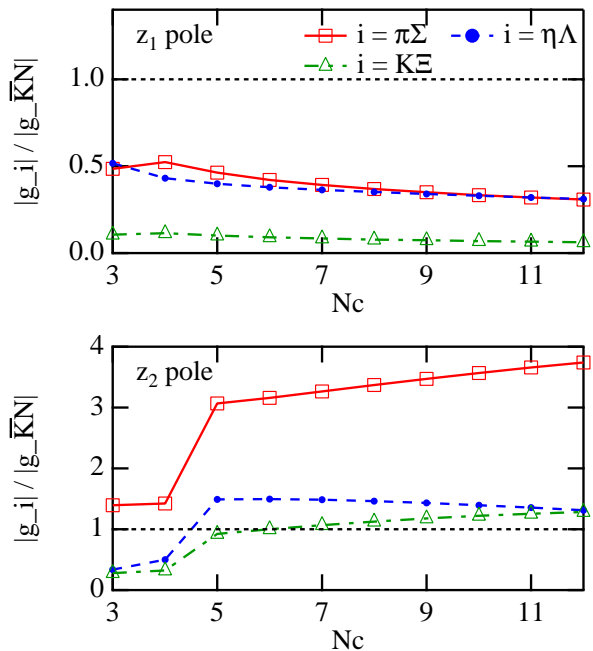}
    \caption{(Color online) Couplings of 
   the poles $z_1$ (upper) and $z_2$ (lower) to the isospin states
   divided by the $\bar K N$ one,
   $|g_i|/|g_{\bar K N}|$.}
    \label{fig:couplings_I}
\end{figure}%

In order to illustrate the properties of the dynamically generated resonances
in large $N_{c}$, it is very interesting to discuss the coupling strengths,
$g_i$, to the different states, which are evaluated by the residues of the
scattering amplitude at the pole positions. In Fig.~\ref{fig:couplings_I}, we
show the couplings of the poles to the different isospin states normalized to
the $\bar K N$ channel for reference: $|g_i|/|g_{\bar{K}N}|$. We can see that
the pole $z_1$ (upper panel), which tends to the bound state, couples 
dominantly to the $\bar{K}N$ state. On the contrary, the pole $z_2$ (lower 
panel), which dissolves into the continuum for large $N_c$, couples 
dominantly to $\pi\Sigma$ and $\bar{K}N$ component becomes less important. 
Analogously, by studying the residues of the poles in the SU(3) basis, we
find that the $z_1$ pole becomes dominantly the flavor singlet state, while 
the $z_2$ pole is dominated by the other components different to the singlet.
These analyses of the coupling strengths indicate that the dominant component
of the pole becoming bound state is flavor singlet ($\bar{K}N$) in the SU(3) 
(isospin) basis, whereas such component in the dissipating resonance becomes 
less important. Hence, we expect that the pole $z_1$ is smoothly connected to
the bound state in the idealized large $N_c$ limit. 

In conclusion, we have addressed the fundamental problem of the structure of 
the $\Lambda(1405)$ baryonic resonance studying the $N_c$ behavior of the two
poles associated to it in the framework of the chiral unitary approach. Based
on the consideration of the standard $N_c$ dependence of the parameters of 
the theory (the meson decay constant and the hadron masses), we have 
discussed important and nontrivial $N_c$ dependence in the Weinberg-Tomozawa
interaction, which leads to an ${{\cal O}(1)}$ attraction in the large $N_c$ 
limit for the flavor singlet and $\bar{K}N$ channels in the SU(3) and isospin
bases, respectively. The attraction in these channels is strong enough to 
create a baryonic bound state. The $N_c$ scaling of the $\Lambda(1405)$ 
resonance shows that one of the two poles tends to become a bound state as 
$N_c$ increases, while the other pole  eventually dissolves into the 
scattering states. The $N_c$ scaling of the decay widths of the poles turns 
out to be at odds with usual QCD predictions for the $N_c$ dependence for 
genuine $qqq$ baryons, indicating that the structures of these poles are not
predominantly $qqq$ states. Thus these findings reinforce the dynamically 
generated nature of the $\Ls$ resonance. Both the results obtained here and 
the methodology used about the $N_c$ behavior of the baryonic resonance go 
one step forward in the understanding of the connection with the underlying 
QCD degrees of freedom and can be used to shed light into the structure of 
other baryonic resonances. 

T.~H. thanks the Japan Society for the Promotion of Science (JSPS) for 
financial support. This work is supported in part by the Grant for Scientific
Research (No.\ 19853500 and No.\ 18042001). L.~R. thanks the following for 
financial support: MEC (Spain) Grants No. FPA2004-03470, No. FIS2006-03438, 
and No. FPA2007-62777; Fundaci\'on S\'eneca Grant No. 02975/PI/05; European 
Union Grant No. RII3-CT-20004-506078; and the Japan (JSPS)-Spain 
collaboration agreement. This research is  part of Yukawa International 
Program for Quark-Hadron Sciences.

\end{document}